\documentclass[final,1p,times,twocolumn,superscriptaddress,preprintnumbers,floatfix,nofootinbib]{elsarticle}
\RequirePackage{lineno}
\usepackage{amssymb}
\usepackage{lipsum}
\usepackage[colorlinks,linkcolor=blue,urlcolor=blue,citecolor=blue]{hyperref}
\usepackage[usenames]{color}

\journal{Nuclear Physics A}

\begin{document}

\begin{frontmatter}



\title{Imaging the structure of atomic nuclei in high-energy nuclear collisions from STAR experiment}

\author[label1,label2]{Chunjian Zhang (for the STAR Collaboration)}
\ead{chunjianzhang@fudan.edu.cn}
\affiliation[label1]{Key Laboratory of Nuclear Physics and Ion-beam Application (MOE), and Institute of Modern Physics, Fudan
University, Shanghai 200433, China}
\affiliation[label2]{Shanghai Research Center for Theoretical Nuclear Physics, NSFC and Fudan University, Shanghai 200438, China}

\begin{abstract}
In relativistic heavy-ion collisions, the extractions of properties of quark-gluon plasma (QGP) are hindered by a limited understanding of its initial conditions, where the nuclear structure of the colliding ions play a significant role. In these proceedings, we present the first quantitative demonstration using ``collective flow assisted nuclear shape imaging" method to extract the quadrupole deformation and triaxiality from $^{238}$U using data from the Relativistic Heavy Ion Collider (RHIC). We achieve this by comparing bulk observables in $^{238}$U+$^{238}$U collisions with nearly spherical $^{197}$Au+$^{197}$Au collisions. A similar comparative measurement performed in collisions of $^{96}$Ru+$^{96}$Ru and $^{96}$Zr+$^{96}$Zr, suggests the presence of moderate quadrupole deformation of $^{96}$Ru, large octupole deformation of $^{96}$Zr, as well as an apparent neutron skin difference between these two species. The prospect of this nuclear shape imaging method as a novel tool for the study of nuclear structure is also elaborated. 
\end{abstract}


\end{frontmatter}




\section{Introduction}
\label{introduction}
The collisions of ultrarelativistic heavy ions generates the QGP, a hot dense phase of nuclear matter, mimicking the first few microseconds after the Big Bang in the early Universe. The expansion and evolution of the system is characterized by the laws of hydrodynamics until about 10 fm/$c$~\citep{Achenbach:2023pba,Chen:2024zwk}. The system then freezes out to form  particles (primarily hadrons) which are observed by the STAR detector. The precise understanding of the initial condition of the QGP are influenced by the nuclear structure of the colliding ions~\citep{Bally:2022vgo}. One can also ask the question if the structure probed at colliders on ultra-short time scales of order $10^{-24} s$ is the same as that measured at low energies. 

Atomic nuclei are composed of $Z$ protons and $N$ neutrons, and their nuclear structure is described by quantum mechanical self-organization~\citep{bohr}. Even the ground states of atomic nuclei show many emergent phenomena including quadrupole, octupole, hexadecapole and triaxial deformations, neutron skin, and nucleonic clusters across the nuclear chart. Their shapes are traditionally measured by spectroscopic techniques at low energies. The nucleon distributions in nuclei are usually modeled by Woods-Saxon (WS) densities, 
\begin{equation}
\rho(r) \propto \left[1+e^{\left[r-R_0\left(1+\beta_2\left(\cos \gamma Y_{2,0}+\sin \gamma Y_{2,2}\right)\right)+\beta_3Y_{3,0}+\beta_4Y_{4,0}\right]/a_0}\right]^{-1}
\end{equation}

\noindent Where $\beta_n$ are the nuclear deformation parameters and $a_0$ is the surface diffuseness parameter. $\gamma$ defines the triaxiality shape, ranging from 0$^{\circ}$ to 60$^{\circ}$. $R_0=1.2 A^{1/3}$ is the nuclear radius. $Y_{n,m}$ are the spherical harmonics.

Collisions of randomly oriented nuclei with a specific structure controls the initial conditions of the QGP, in particular enhancing the fluctuations of its ellipticity and overlap area in the transverse ($xy$) plane. They are quantified by $\varepsilon_2=\left(\left\langle y^2\right\rangle-\left\langle x^2\right\rangle\right)/\left(\left\langle y^2\right\rangle+\left\langle x^2\right\rangle\right)$ and $R^{2}_{\perp} = 1/d_{\perp} \propto 1 / \sqrt{\left\langle x^2\right\rangle\left\langle y^2\right\rangle}$ from nucleon distributions, respectively. The final state $v_2$ and event-averaged transverse momentum $\delta p_{\mathrm{T}}$ indeed emerge as a response to the initial state, $v_2 \propto \varepsilon_2$ and $\delta p_{\mathrm{T}} \propto \delta d_{\perp}$ based on hydrodynamics~\citep{Giacalone:2021udy}.

In these proceedings, we will address the question of whether the values of nuclear parameters found in the low-energy literature are consistent with experimental data at high energies. We present observational evidence for this, and argue that, at present, the study of consistency of nuclear experiments across energy scales is a crucial interdisciplinary research area at the interface of nuclear physics and high-energy physics.

\section{Extractions of quadrupole and triaxiality deformation of $^{238}$U}
As shown in~\citep{NN2024c} and~\citep{STAR:2024eky}, the signals of $\left\langle v_2^2\right\rangle$ and $\left\langle\left(\delta p_T\right)^2\right\rangle$ in most central U+U are strongly enhanced compared to that in Au+Au collisions. These features demonstrate the geometric role of large $\beta_{\rm{2,U}}$~\citep{Jia:2021tzt,Giacalone:2023hwk}. These two systems have nearly equal mass number (A) but completely different nuclear shapes, with $^{197}$Au nuclei having a slightly oblate shape~\citep{Bally:2023dxi}. As a result, taking ratios between these two collision systems almost completely cancels out final state effects, leaving model uncertainties primarily due to initial conditions. We have found a simple parametric dependence on shape parameters, $\left\langle v_2^2\right\rangle=a_1+b_1 \beta_2^2$ and $\left\langle\left(\delta p_{\mathrm{T}}\right)^2\right\rangle=a_2+b_2 \beta_2^2$ in model simulations. The coefficient $a$ is in the absence of deformation and minimized in central collisions, while $b$ quantifies how efficiently fluctuations in the global geometry translate into deformation~\citep{Giacalone:2021udy}.

The sign change of the covariance $\left\langle v_2^2 \delta p_{\mathrm{T}}\right\rangle$ in U+U is observed in central collisions, while the results from Au+Au remain positive throughout the centralities. This strong suppression is expected for a large $\beta_2$ of $^{238}$U as described in the parametric form, $\left\langle v_2^2 \delta p_{\mathrm{T}}\right\rangle=a_3-b_3 \beta_2^3 \cos (3 \gamma)$~\cite{Jia:2021qyu,Dimri:2023wup}. 

The event-averaged moments of these observables capture the two- and three-body nucleon distributions in the intrinsic frame that are most predominant in ultra-central collisions. The ratios between U+U at $\sqrt{s_{NN}}=$ 193 GeV and Au+Au at $\sqrt{s_{NN}}=$ 200 GeV collisions in the 0–5\% most central region, have the greatest sensitivity to the $^{238}$U shape. By comparison with several state-of-the-art hydrodyamical models, IP-Glasma+MUSIC+UrQMD~\cite{Schenke:2020mbo} and Trajectum~\cite{Nijs:2023yab}, these measurements provide capability for the first quantitative and simultaneous determination of $\beta_2$ and $\gamma$. The comparisons between the data and the IP-Glasma+MUSIC+UrQMD model favor $\beta_{2\rm{U}}$ ranges,  
$\beta_{2 \rm{U}}=0.297 \pm 0.015$ and $\gamma_{\rm{U}}=8.5^{\circ} \pm 4.8^{\circ}$ with a combined analysis of constraints from the ratios $R_{\left(\delta p_{\mathrm{T}}\right)^2}$ and $R_{v_2^2 \delta p_{\mathrm{T}}}$. Note that Trajectum with a different implementation of the initial condition and QGP evolution, are only tuned based on Bayesian analysis of the LHC data. We have extrapolated to the RHIC energies in the current analysis. Nevertheless, it is quite useful to combine two different hydrodynamic models together in order to estimate the theoretical uncertainties yielding $\beta_{2 \rm{U}}=0.286 \pm 0.025$ and $\gamma_{\rm{U}}=8.7^{\circ} \pm 4.5^{\circ}$. 

The results are remarkably consistent with the low energy measurement, confirming that the $^{238}$U are largely deformed~\cite{Pritychenko:2013gwa}. Meanwhile, the small $\gamma_{\rm U}$ value demonstrates the first extraction of nuclear ground state triaxiality without involving transitions to excited states.

\section{Nuclear structure in $^{96}$Ru and $^{96}$Zr nuclei}
\begin{figure}[hbtp]
	\centering 
	\includegraphics[width=0.7 \textwidth, angle=0]{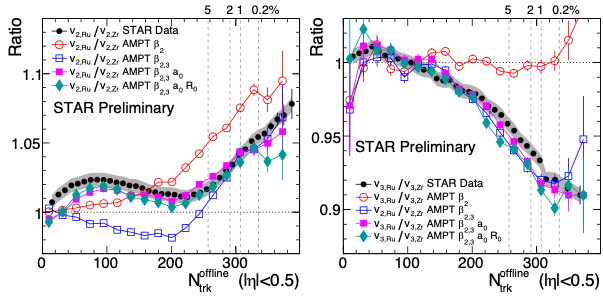}	
	\caption{The ratios of $v_2$ (left panel) and $v_3$ (right panel) between $^{96}$Ru+$^{96}$Ru and $^{96}$Zr+$^{96}$Zr collisions as a function of $\rm{N_{trk}^{offline}}$ at $\sqrt{s_{NN}}=$ 200 GeV. The results of AMPT models with different nuclear parameters were compared.} 
	\label{fig1}%
\end{figure}

Since isobar nuclei have the same mass number, $A$ but different structures, the final state effects are canceled by taking ratios~\citep{Zhang:2022fou}. Therefore, any deviation from unity in the ratio of any bulk observable must be due to differences in the structure of the colliding nuclei, that would affect the initial conditions of the QGP~\cite{Zhang:2021kxj,Jia:2022qgl}. Figure~\ref{fig1} shows the ratios of $v_2$ (left panel) and $v_3$ (right panel) between $^{96}$Ru+$^{96}$Ru and $^{96}$Zr+$^{96}$Zr collisions as a function of the charged hadron multiplicity with $|\eta|<$0.5 at $\sqrt{s_{NN}}=$ 200 GeV. In particular, the characteristic broad peak and non-monotonic behavior of the $v_2$ ratio in mid-central and peripheral collisions is a clear signature of the influence of the neutron skin difference $\Delta a$ between these two species~\cite{Xu:2022ikx,Jia:2021oyt}. In near central collisions, this ratio is influenced by a positive contribution from $\Delta \beta_2^2$ and a larger negative contribution from $\Delta \beta_3^2$. The enhancement of $v_2$ in the central region originates from the quadrupole deformation $\beta_2$ of $^{96}$Ru, while the intriguing trends of $v_3$ are mainly influenced by the octupole deformation $\beta_3^2$ of $^{96}$Zr.

As shown in phenomenological studies, the ratio of observable $\mathcal{O}$ between isobar-like or isobar collisions has a simple scaling relation, 
\begin{equation}
R_{\mathcal{O}} \equiv \frac{\mathcal{O}_{\mathrm{Ru}}}{\mathcal{O}_{\mathrm{Zr}}} \approx 1+c_1 \Delta \beta_2^2+c_2 \Delta \beta_3^2+c_3 \Delta R_0+c_4 \Delta a
\end{equation}
where $\Delta \beta_n^2=\beta_{n, \mathrm{Ru}}^2-\beta_{n, \mathrm{Zr}}^2, \Delta R_0=R_{0, \mathrm{Ru}}-R_{0, \mathrm{Zr}}, \Delta a=$ $a_{\mathrm{Ru}}-a_{\mathrm{Zr}}$ and $c_n=b_n / b_0$. These ratios can probe the difference in the WS parameters between the isobar nuclei, and the contributions among the WS parameters are independent of each other  We simulated collision events using a multi-phase transport (AMPT)~\cite{Lin:2004en} model varying the nuclear structure parameters to match the STAR data. The extracted quadrupole deformation value for $^{96}$Ru is $\beta_{2, \rm{Ru}}=0.16 \pm 0.02$ and the octupole deformation value for $^{96}$Zr is $\beta_{3,\rm{Zr}}=0.20 \pm 0.02$. Here, the reported uncertainty combines statistical and systematic uncertainties. In addition, we also constrain the nuclear size and neutron skin differences to $\Delta R_0=$ 0.07 $fm$ and $\Delta a=$ -0.06 $fm$, respectively. Note that the actual fine radial structures, as described by density functional theory based on effective nuclear forces, have only a few percent uncertainties in the measured differences in mid-central collisions, which can be absorbed as model systematic uncertainties~\citep{Yan:2024ggt}.

\section{Summary and outlook}
In summary, we have quantitatively constrained the nuclear deformation parameters $\beta_2$ and $\gamma$ of the $^{238}$U nucleus by comparing the nearly spherical $^{197}$Au nucleus simultaneously using three bulk observables. Our extractions are remarkably consistent with the low energy estimate based on a rigid-rotor assumption. We have also extracted $\beta_2$ for $^{96}$Ru, $\beta_3$ for $^{96}$Zr, and the difference in neutron skin $\Delta a$ between these two species. This innovative shape imaging approach consequently helps to improve the QGP initial state parametrization and is a conducive step to facilitate the interdisciplinary research between low nuclear and high energy physics.

The fluctuations in the initial state extend beyond the $xy$ plane and are also evident in the longitudinal direction. Currently, quantitative evaluations of longitudinal fluctuations are still limited by rapidity coverage and lack of data. Comparing collisions of atomic nuclei with similar $A$ but very different shapes could provide a deep understanding of the longitudinal decorrelations by investigating the rapidity dependence of bulk observables and their differences~\citep{Zhang:2024bcb,Jia:2024xvl}. 

It is also crucial to benchmark the shape imaging method to the structure of light nuclei, where the mean-field description breaks down and nucleon-nucleon correlations gain importance primarily described by the modern $ab$ $initio$ approaches~\citep{Elhatisari:2022zrb,Huang:2023viw,Giacalone:2024luz,Zhang:2024vkh,Chen:2018tnh,Noronha:2024dtq}. Combining present and future analyses, we expect that the method of nuclear shape imaging in high-energy nuclear collisions will offer a high-resolution view of the initial conditions of QGP and a deep understanding of fundamental questions of nuclear structure.

\bibliographystyle{elsarticle-num}
\bibliography{references}{}






\end{document}